\documentclass[reprint,
superscriptaddress,
amsmath,amssymb,aps,showkeys,showpacs,
twoside,final,secnumarabic,%raggedbottom,
nofootinbib]{revtex4-2}

\usepackage[paperwidth=205mm,paperheight=290mm,top=17mm,bottom=25mm,
inner=17mm,outer=17mm,
twoside]{geometry}

\usepackage{cmap} % Улучшенный поиск %русских слов в~полученном pdf-файле
\usepackage[T1,T2A]{fontenc}
\usepackage[utf8]{inputenc}
\usepackage[russian,english]{babel}
\usepackage{color}
\usepackage{graphicx}% Include figure files
\usepackage{dcolumn}% Align table columns on decimal point
\usepackage{bm} % bold math
\usepackage[unicode=true,colorlinks=true,linkcolor=magenta, urlcolor=blue, citecolor = blue,breaklinks]{hyperref}
\usepackage{multirow}
\usepackage{url}
\usepackage{breakurl}
\usepackage{subcaption}
\usepackage{comment}
\usepackage{caption}

\DeclareGraphicsExtensions{.eps}

\newcount\issue
\newcount\Vol
\newcount\numb
\usepackage{fancyhdr} %this packages %provides fancy up and bottom of page
\def\Vol{\textbf{78}}
\def\numb{x}
\usepackage{enumitem}
\usepackage[export]{adjustbox}
\usepackage{tikz}

\begin{document}

%====== Начало шапки статьи  ============
\title
%{22nd Lomonosov Conference on Elementary Particle Physics\\[20pt]
{New effect in neutrino spin oscillations in transversal matter currents
\\with nonstandard interactions}

\def\addressa{Department of Physics, Branch
of Lomonosov
Moscow State University in Sarov, 607328 Nizhny Novgorod Region, Russia, }
\def\addressb{Department of Theoretical Physics \& Department of Particle Physics and Extreme States of Matter, Faculty of Physics, Moscow State University, 119992 Moscow, Russia}

\author{\firstname{Inna}~\surname{Kozlovskaya}}
% \email[E-mail: ]{KolovskayaIS@my.msu.ru}
\affiliation{\addressa} \affiliation{\addressb}
\author{\firstname{Alexander}~\surname{Studenikin}}
\affiliation{\addressa} \affiliation{\addressb}

%\affiliation{\addressa}
%\affiliation{\addressb}

%\received{xx.xx.2023}
%\revised{xx.xx.2023}
%\accepted{xx.xx.2023}

\begin{abstract}

We perform a systematic study of a new phenomenon of neutrino spin and spin‑flavor oscillations induced by transverse matter currents, based on the developed quantum treatment of the phenomenon. Both standard and nonstandard interactions of neutrinos with the external medium are taken into account. As an example, the oscillations in question are considered under the conditions of a binary neutron star merger model.
\end{abstract}

%DOI:

\maketitle
\thispagestyle{fancy}

%====== Начало  статьи  ============

\section{Introduction}\label{intro}
For about a century, neutrinos have offered a unique probe of physics beyond the Standard Model. Within the Standard Model itself, neutrinos are predicted to be massless; however, the discovery of neutrino flavor oscillations - a process that requires finite mass - demonstrates the need for an extended theoretical framework. Among the possible new-physics effects, spin oscillations driven by interactions with a moving external background \cite{key1} are especially worthy of attention.  Directed streams of matter are
often found in the vicinity of extreme objects, so the
latter create suitable conditions for searching for the
effects of new physics. \\

In astrophysical conditions, neutrino oscillations can be strongly affected by magnetic fields and matter and they can trigger both flavor and spin oscillations. In the presence of magnetic fields and matter the three-dimensional neutrino spin vector evolves according to the Bargmann-Michel-Telegdi equation \cite{key2}\\

\begin{equation}
\frac{d\mathbf{S}}{dt} = \frac{2\mu}{\gamma} \left[ \mathbf{S} \times (\mathbf{B}_0 + \mathbf{M}_0) \right],
\end{equation}
 where the magnetic field $\mathbf{B}_0$ and the matter term $\mathbf{M}_0$ in the neutrino rest frame are determined by the corresponding longitudinal and transverse fields components given in the laboratory frame:

\begin{equation}
\mathbf{M}_0 = \mathbf{M}_{\parallel} + \mathbf{M}_{\perp},
\end{equation}

\begin{equation}
\mathbf{M}_{\parallel} = \frac{G_F(1 + 4 \sin^2 \theta_W)}{2\sqrt{2}\mu\sqrt{1 - v_e^2}} n_0 \left( 1 - \frac{{\bm v}_e {\bm \beta}}{1 - \gamma^{-2}} \right) \gamma \beta,
\end{equation}

\begin{equation}
\mathbf{M}_{\perp} = -\frac{G_F(1 + 4 \sin^2 \theta_W)}{2\sqrt{2}\mu\sqrt{1 - v_e^2}} n_0 \mathbf{v}_{e_\perp}.
\end{equation}
Here $\mu$ is the neutrino magnetic moment, ${\bm \beta}$ is the neutrino velocity and $\gamma = (1-\beta ^2)^{-\frac{1}{2}}$.

 Using the quasiclassical equation for describing the evolution of the neutrino spin in a magnetic field and in matter, and retaining the first-order contributions $O(\gamma^{-1})$ in the expansion of the matter potential $M_0$ in terms of the neutrino Lorentz factor $\gamma$, it was shown for the first time in \cite{key1} that spin and spin-flavor oscillations arise in the case of neutrino propagation in moving matter in the presence of a matter current transverse to the neutrino direction of motion (or transverse matter polarization).
 The paper \cite{key5} also provides a consistent quantum description of this phenomenon.

In this work, the total probabilities of oscillations under the influence of longitudinal and transversal currents of the matter are considered. Explicit expressions are obtained for the probabilities of oscillations under the influence of medium currents and nonstandard interactions (NSI), and numerical values are obtained for the probabilities of oscillations under the conditions of a merger of two neutron stars for both cases.
\newline

\section{\label{sec:level1}The equations of evolution}

Here, we study the effect of
 neutrino spin evolution induced by transversal matter
 currents and use a consistent derivation (see also
 \cite{key5}) of the effect based on the direct calculation of
 the spin evolution effective Hamiltonian for a
 neutrino propagating in transversal currents of matter. Consider two flavor neutrinos with two possible helicities $\nu_f = (\nu_e^+, \nu_e^-, \nu_{\mu}^+, \nu_{\mu}^-)^T$ in moving matter composed
 of neutrons. The neutrino interaction Lagrangian reads:

\begin{eqnarray}
    L_{int}&&= -f^{\mu}\sum_l \bar\nu_l(x)\gamma_{\mu}\frac{1+\gamma_5}{2}\nu_l(x) \nonumber\\
 && =  -f^{\mu}\sum_i \bar\nu_i(x)\gamma_{\mu}\frac{1+\gamma_5}{2}\nu_i(x), \\
    f^{\mu} &&= -\frac{G_F}{\sqrt{2}}n(1,\mathbf{v}) \nonumber,
\end{eqnarray}
where $l = e, \mu$ , $i = 1,2$  and the matter potential $f^{\mu}$ depends on the neutron number density in the laboratory reference frame $n = \frac{n_0}{\sqrt{1-v^2}}$  and on the velocity of matter $\mathbf{v} = (v_1, v_2, v_3)$. The neutrino evolution equation in the flavor basis is
 \begin{equation}
     i\frac{d}{dt}\nu_f=(H_0 + \Delta H_{SM})\nu_f,
 \end{equation}
 where $H_0$ is the effective Hamiltonian that determines the neutrino evolution in non-moving matter and $\Delta H_{SM}$ accounts for the effect of motion of the matter. The evolution equation allows one to obtain the square of the matrix element, which gives the probability of the corresponding oscillations.\\

 To account for nonstandard interactions, consider a Lagrangian of the form \cite{keyFarzan}
\begin{equation}
    -L^{eff}_{NSI}=\epsilon_{\alpha\beta}^{fP}2\sqrt{2}G_F(\bar\nu_\alpha\gamma_\rho L\nu_\beta)(\bar f \gamma^\rho Pf),
\end{equation}
\begin{equation}
 L, R = \frac{(1 \pm \gamma^5)}{2},
\end{equation}
 where $f = u, d, e$ and $\alpha, \beta = e, \mu$.
In this case the neutrino evolution equation with standard and nonstandard interactions in the flavor basis is
\begin{equation}\label{ham}
    i\frac{d}{dt}\nu_f=(H_0 + \Delta H_{SM} + \Delta H_{NSI})\nu_f.
\end{equation}
The corresponding oscillation probabilities are presented below.
 \newline

\section{Numerical Model for a neutrino propagation in a binary neutron star merger}
A binary neutron star
 merger forms initially a central,
 hypermassive neutron star (HMNS) surrounded by a thick
 accretion disc. During the merger process a small fraction
 of the total mass becomes ejected via gravitational torques
 and hydrodynamic processes \cite{key3}. For the purposes of further estimation, we adopt the following values of the specific parameters describing the neutrinos and the environmental medium in question:

\begin{itemize}
\setlength{\itemsep}{-0.1ex} % Уменьшает расстояние между пунктами
\setlength{\parskip}{7pt}
\item the transversal velocity of the medium $v_\perp = 0.067$,
\item the neutrino energy $p_0 = 10^6 eV$,
\item the background neutron concentration $n = 5 \times 10^{36} cm^{-3}$,
\item the sine of the mixing angle $\sin^2\theta = 0.297$,
\item the neutrino mass squares difference $\Delta m^2_{eff} \approx 10^{-4} eV^2$,
\item the neutrino masses are used  $m_1 = 0 , m_2 = 10^{-2} eV$,
\item the NSI parameters are set to be equal $\epsilon_{ee} = 0.17 $, $\epsilon_{\mu\mu} = 1.5 $, $\epsilon_{e\mu} = 0.84 $.
\end{itemize}
It is considered that the largest neutrino flux is formed in the polar directions $\phi = 0, \phi = \pi$.
For the neutrino–quark NSI couplings, we adopt the upper limits obtained from the constraints given in \cite{key4}.

\vspace{15pt}
\begin{figure}
\center
 \includegraphics[height=5cm]{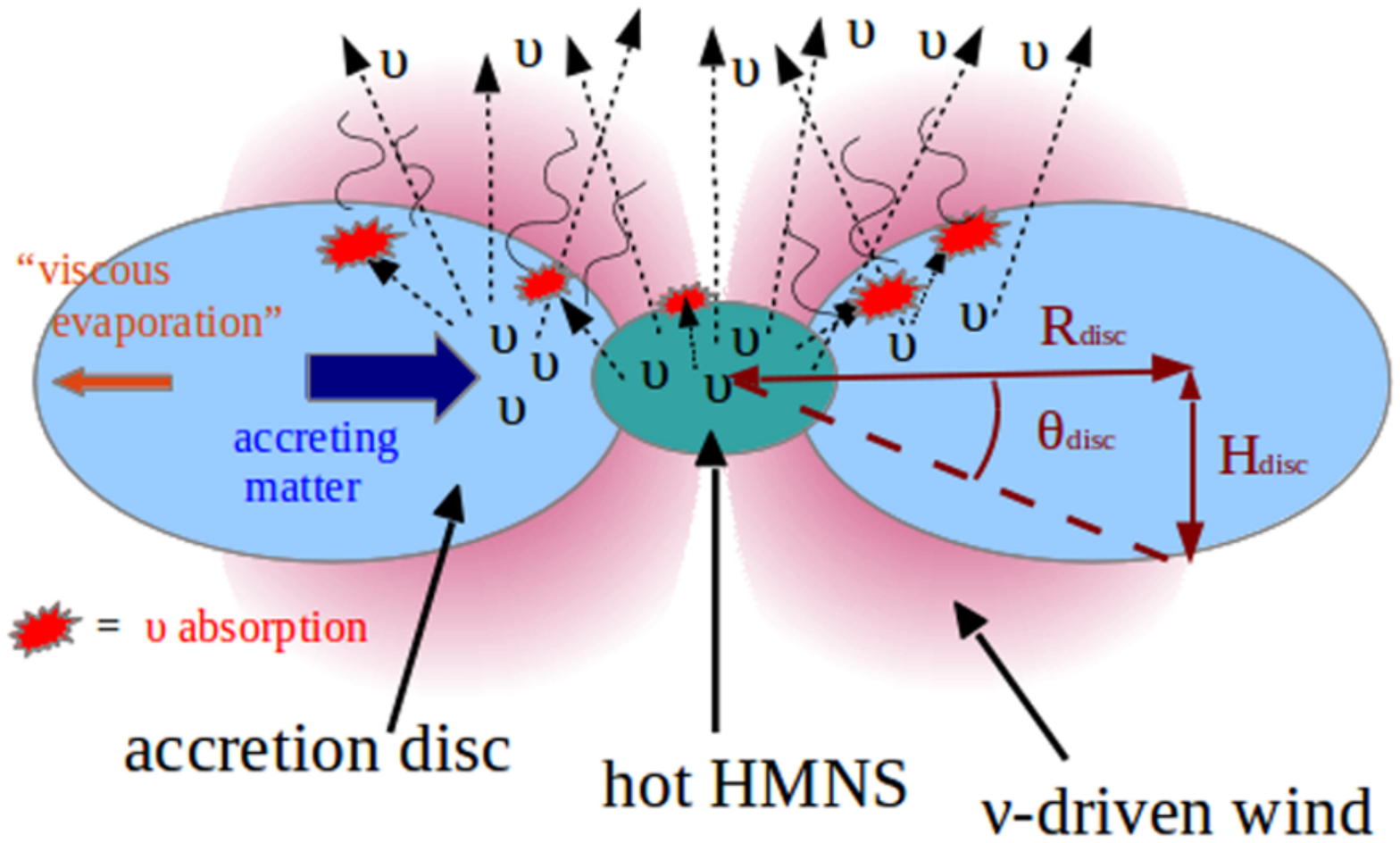}
 \\{\textbf{Fig.1.} A model of binary neutron star merger presented in \cite{key3}.}
\end{figure}
%\vspace{-10 pt}
\section{The probability of oscillations under the influence of matter currents}
%\vspace{-20 pt}

Using the standard calculation scheme based on the exact expressions for the effective neutrino evolution Hamiltonian (9), one can obtain the following expression for the probability of neutrino spin oscillations $\nu_e^L \rightarrow \nu_e^R$  in a transverse matter current $j_\perp$

\begin{equation}
    P_{\nu_e^L \rightarrow \nu_e^R}^{(j_\perp)}(t) = \frac{\left(\frac{\eta}{\gamma}\right)^2 _{ee} v_\perp^2}{\left(\frac{\eta}{\gamma}\right)^2 _{ee} v_\perp^2 + (1 - \mathbf{v}{\bm \beta})^2} \sin^2 \omega_{ee}^{j_\perp} t
    \label{eq:10}.
\end{equation}
The probability as a function of time $t$ (or neutrino travel distance $x$) is shown in Fig.2.
In a similar way,  the probability $P^{j_\perp}_{\nu_e^L \rightarrow \nu_\mu^R}(t) $ for the neutrino spin-flavour oscillations $\nu_e^L \rightarrow \nu_\mu^R$ in a transversal matter current $j_\perp$ can be obtained .

Using the expression for the oscillation probability in the case of longitudinal motion of the medium relative to the neutrino propagation direction \cite{Grigoriev:2002ew}, and following the analogy with the case \cite{Popov:2019nkr}  of neutrino oscillations in the presence of a magnetic field, for full probability of oscillations one can get (see also  \cite{key5,key6})

%\vspace{-39 pt}
 \begin{equation}
    P_{\nu_e^L \rightarrow \nu_\mu^R}^{(j_\parallel + j_\perp)}(t) = \left(1 - P_{\nu_e^L \rightarrow \nu_e^R}^{(j_\perp)} - P_{\nu_e^L \rightarrow \nu_\mu^R}^{(j_\perp)}\right) P_{\nu_e^L \rightarrow \nu_\mu^L}^{(j_\parallel)}.
\label{eq:11}
\end{equation}
The probability as a function of time $t$ (or neutrino travel distance $x$) is shown in Fig.3.

\begin{figure}[h]
\centering

\includegraphics[width=0.73\columnwidth]{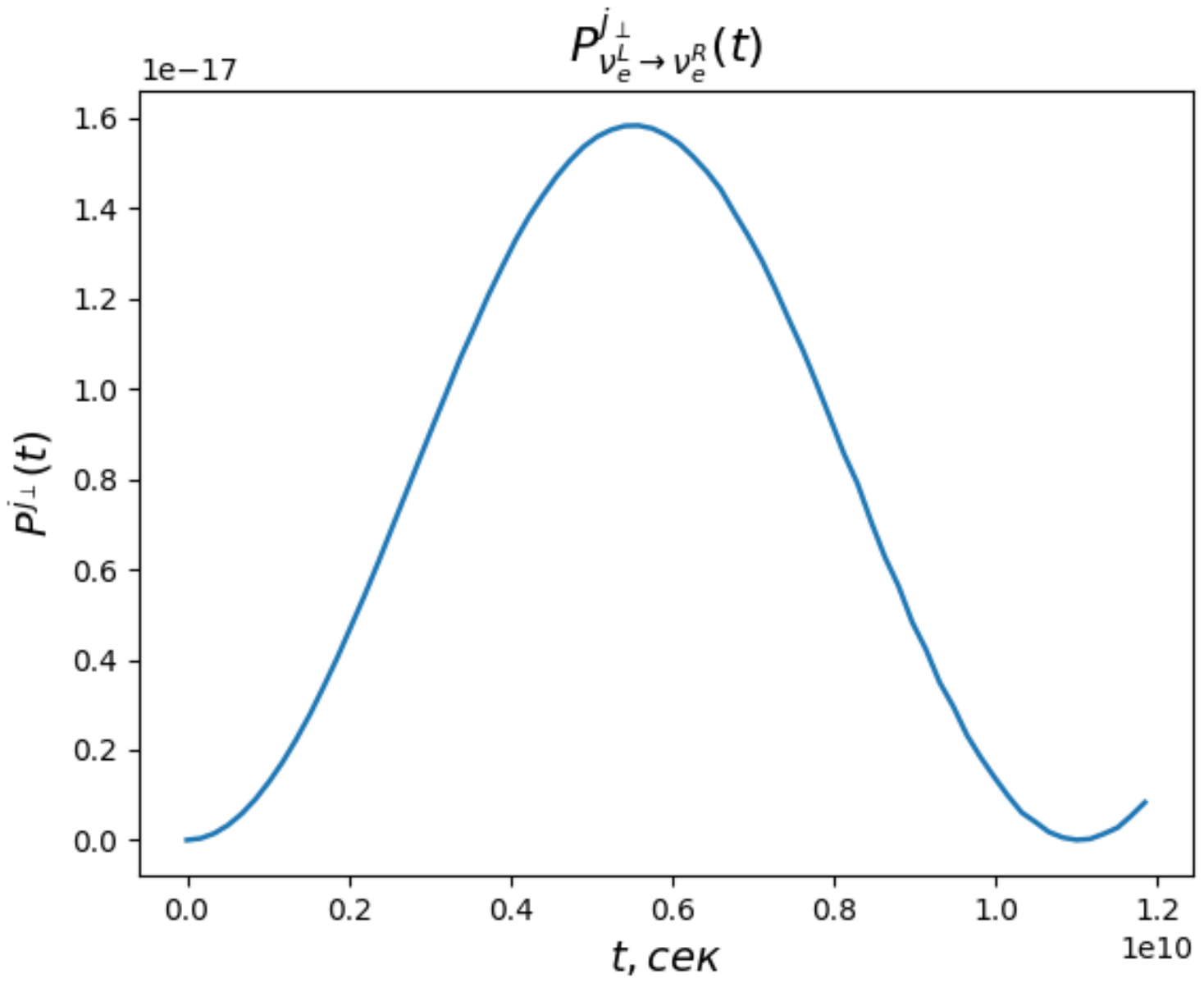}
\\ \vspace{0.2cm}
\textbf{Fig.2.} The probability of neutrino spin oscillations $\nu_e^L \rightarrow \nu_e^R$  in a transverse matter current $j_\perp$.
\label{fig:1.png}

\vspace{0.3cm}

\includegraphics[width=0.73\columnwidth]{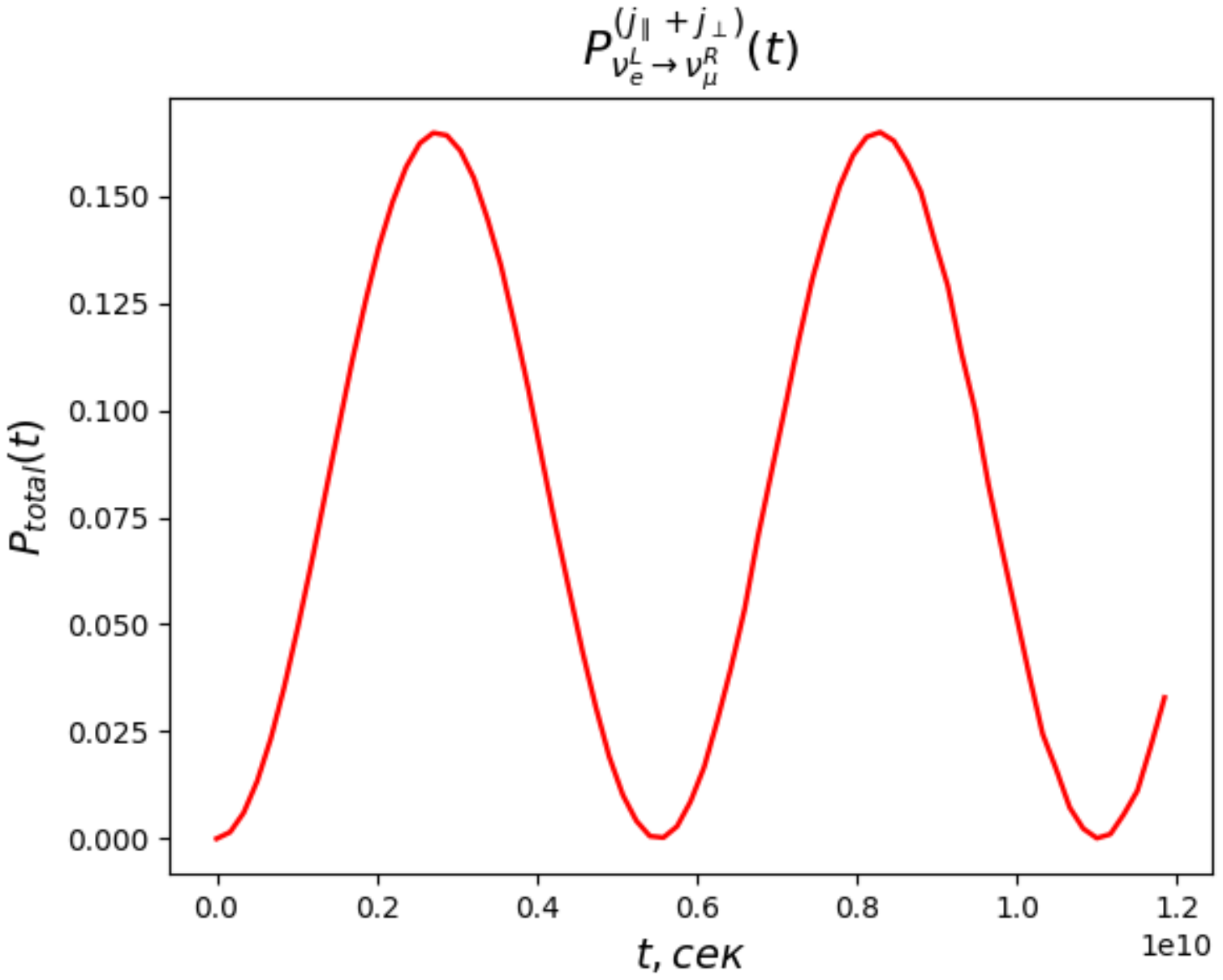}
\\ \vspace{0.2cm}
%\raggedright
\textbf{Fig.3. } The full probability of neutrino spin-flavour oscillations  $\nu_e^L \rightarrow \nu_\mu^R$ in arbitrary moving matter.
\label{fig:4.png}

%\caption{The numerical solution for the probability of oscillations under the influence of matter currents.}
\label{fig:epsart}
\end{figure}
 The following notation is used here:

%\newpage
\begin{align}
 \left(\frac{\eta}{\gamma}\right)_{ee} &= \frac{\cos^2\theta}{\gamma_{11}}+\frac{\sin^2\theta}{\gamma_{22}},\\
\gamma_{\alpha\alpha'}^{-1}&=\frac{1}{2}(\gamma_\alpha^{-1}+\gamma_{\alpha'}^{-1}), \\
\gamma_\alpha^{-1} &= \frac{m_\alpha}{E_\alpha}, \alpha = 1, 2, \\
\omega_{e e}^{j_\perp} &= \tilde{G} n \sqrt{\left(\frac{\eta}{\gamma}\right)^2_{ee}  v_\perp^2 + (1 - \mathbf{v}{\bm \beta})^2}.
\label{eq:pythagoras}
\end{align}

\begin{figure}[h]
\centering

\includegraphics[width=0.73\columnwidth]{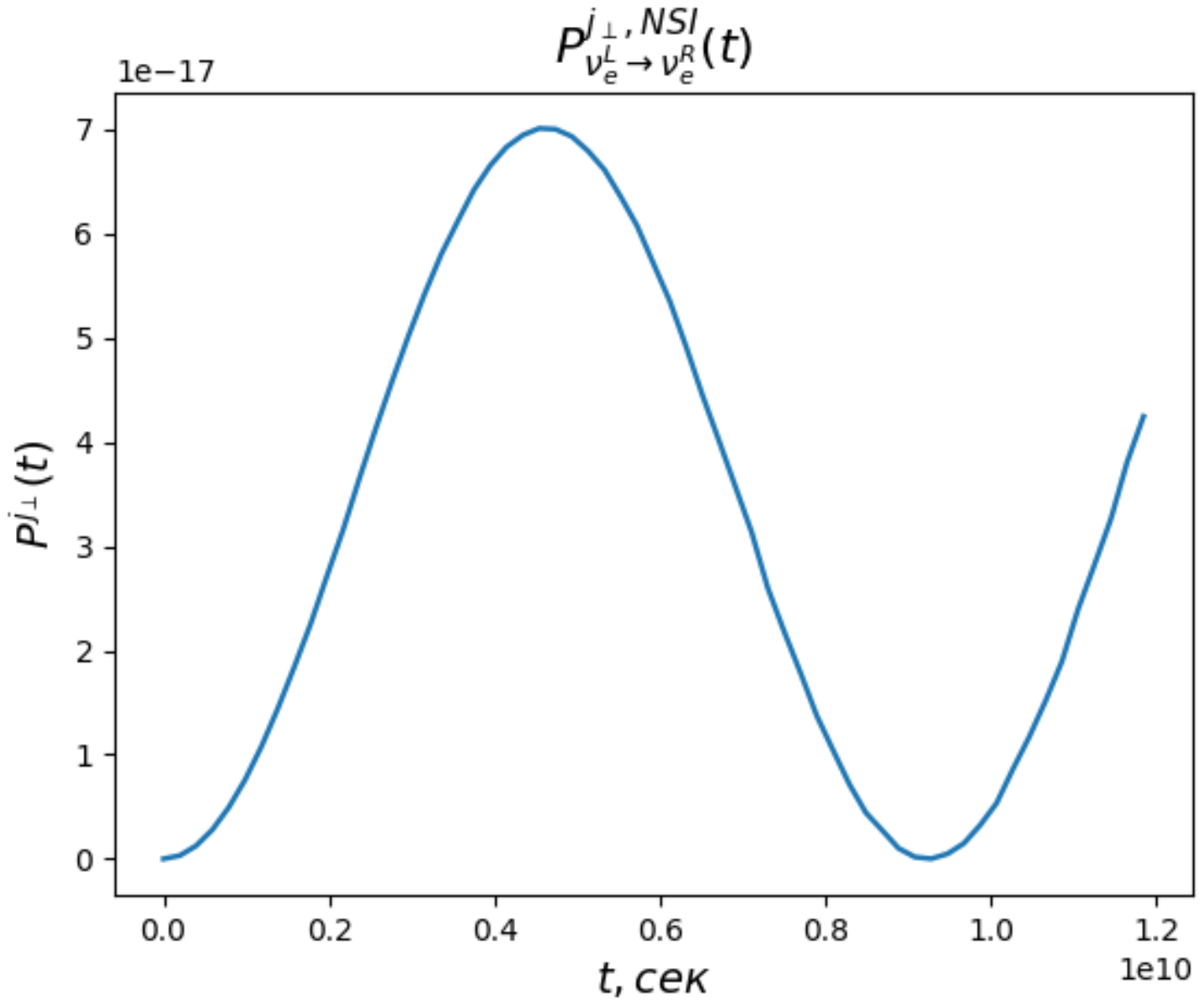}
\\ \vspace{0.2cm}
\textbf{Fig.4.} The probability of neutrino spin oscillation $\nu_e^L \rightarrow \nu_e^R$ under the influence of a transversal current with NSI.
\label{fig:3.png}
\end{figure}

\begin{figure}[h]
\includegraphics[width=0.73\columnwidth]{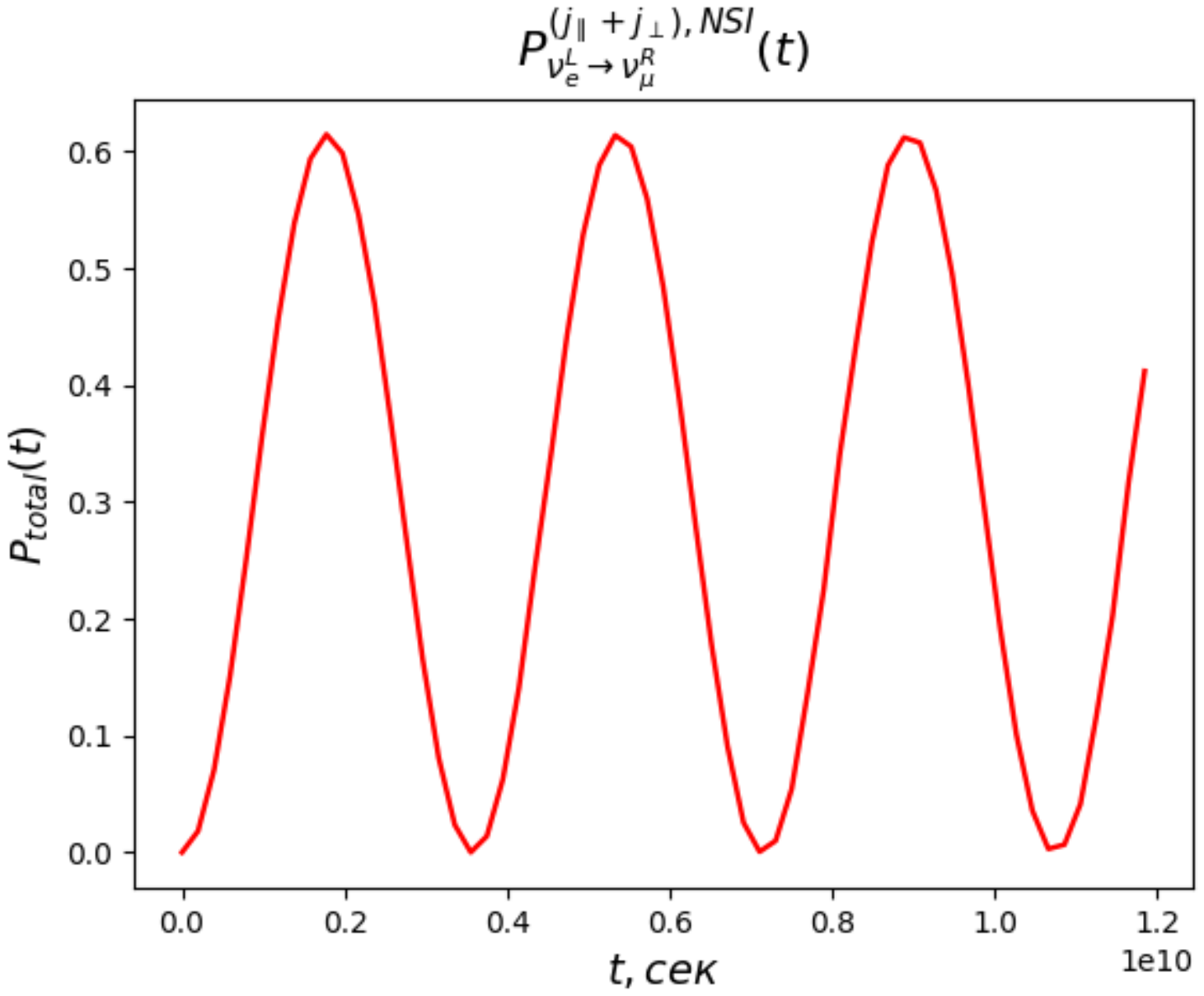}
\\ \vspace{0.2cm}
\textbf{Fig.5.} The full  probability of neutrino spin-flavour oscillations $\nu_e^L \rightarrow \nu_\mu^R$ in arbitrary moving matter with NSI.
\label{fig:4.png}
\end{figure}

%\twocolumngrid

\section{The probability of oscillations under the influence of matter currents with NSI}

%\begin{figure}[b]
%\centering

%\includegraphics[width=0.8\columnwidth]{5.png}
%\\ \vspace{0.2cm}
%\textbf{(a)} Probability of spin oscillation under the influence of transversal current with NSI
%\label{fig:5.png}

%\vspace{0.3cm}

%\includegraphics[width=0.8\columnwidth]{8.png}
%\\ \vspace{0.2cm}
%\raggedright
%\textbf{(b)} Total probability of spin-flavour oscillations with NSI
%\label{fig:8.png}

%\caption{Numerical solution for probability of oscillations under the influence of matter currents with NSI}
%\label{fig:epsart}
%\end{figure}

The developed approach enable us to consider the case of the neutrino propagation under the influence of the transversal matter current also with accounting for possible NSI.
For the probability of neutrino spin oscillations ${\nu^L_e \xrightarrow{} \nu_e^R}$  in this case we get

\begin{widetext}
\begin{equation}
 P_{\nu^L_e \xrightarrow{} \nu_e^R}^{j_\perp, NSI} (t)= \frac{\left(v_\perp \left[\frac{1+\epsilon_{ee}}{2\gamma_{21}}+\frac{1+\epsilon_{ee}}{2}\left(\left(\frac{\eta}{\gamma}\right)_{ee} - \left(\frac{\eta}{\gamma}\right)_{\mu\mu}\right) + \left(\frac{\eta}{\gamma}\right)_{e\mu}\right]\right)^2}{\left(v_\perp \left[\frac{1+\epsilon_{ee}}{2\gamma_{21}}+\frac{1+\epsilon_{ee}}{2}\left(\left(\frac{\eta}{\gamma}\right)_{ee} - \left(\frac{\eta}{\gamma}\right)_{\mu\mu}\right) + \left(\frac{\eta}{\gamma}\right)_{e\mu}\right]\right)^2 + ((1+\epsilon_{ee})(1-v_\parallel))^2}\sin^2(\omega_{s} t),
\end{equation}
 \label{eq:wideeq}
%\end{widetext}
where
%\begin{widetext}
\begin{equation}
\omega_{s} = \tilde{G}n\sqrt{\left(v_\perp \left[\frac{1+\epsilon_{ee}}{2\gamma_{21}}+\frac{1+\epsilon_{ee}}{2}\left(\left(\frac{\eta}{\gamma}\right)_{ee} - \left(\frac{\eta}{\gamma}\right)_{\mu\mu}\right) + \left(\frac{\eta}{\gamma}\right)_{e\mu}\right]\right)^2 + \left((1+\epsilon_{ee})(1-v_\parallel)\right)^2}.
\end{equation}
\end{widetext}

The following notations are used:
\begin{equation}
  \left(\frac{\eta}{\gamma}\right)_{\mu\mu}=\frac{\sin^2\theta}{\gamma_{11}}+\frac{\cos^2\theta}{\gamma_{22}},
\end{equation}

\begin{equation}
 \left(\frac{\eta}{\gamma}\right)_{e\mu} = \frac{\sin2\theta}{\tilde{\gamma}_{21}},
\end{equation}

\begin{equation}
\tilde{\gamma}_{21}^{-1} =\frac{1}{2}(\gamma_2^{-1}-\gamma_1^{-1}).
\end{equation}

The corresponding oscillation probability is shown in Fig.4.

For the full neutrino spin-flavour oscillations $\nu_e^L \rightarrow \nu_\mu^R$ probability in an arbitrary moving matter with standard and nonstandard interactions we get (see Fig.5)

\begin{equation} \scriptsize
    P_{\nu_e^L \rightarrow \nu_\mu^R}^{(j_\parallel + j_\perp), NSI}(t) = \left(1 - P_{\nu_e^L \rightarrow \nu_e^R}^{(j_\perp), NSI} - P_{\nu_e^L \rightarrow \nu_\mu^R}^{(j_\perp), NSI}\right) P_{\nu_e^L \rightarrow \nu_\mu^L}^{(j_\parallel), NSI}.
\end{equation}
The exact expressions for the oscillations probabilities $P^{j_\perp, NSI}_{\nu_e^L \rightarrow \nu_\mu^R}$ and $P^{j_\parallel, NSI}_{\nu_e^L \rightarrow \nu_\mu^L}$ can be straightforwardly derived as generalization of the corresponding probabilities  $P^{j_\perp}_{\nu_e^L \rightarrow \nu_\mu^R}$ and $P^{j_\parallel}_{\nu_e^L \rightarrow \nu_\mu^L}$ in an arbitrary moving matter without NSI.

It is noteworthy that the amplitudes of neutrino oscillations including NSI (non-standard neutrino interactions) are greater than those without NSI.

\section{Summary}
In this study, we employed a quantum-mechanical approach to analyze neutrino spin $\nu^L_e \xleftarrow{} (j_\perp) \rightarrow \nu_e^R$ and spin-flavor $\nu^L_e \xleftarrow{} (j_\perp) \rightarrow \nu_\mu^R$ oscillations induced \newpage\noindent by transversal matter currents. This phenomenon was originally predicted in \cite{key1} based  on a semiclassical treatment of neutrino spin evolution in matter.  For definiteness, we considered a neutron-dominated medium.

In addition, we have presented a detailed theoretical study of neutrino spin oscillations $\nu^L_e \xleftarrow{} (j_\perp) \rightarrow \nu_e^R$ engendered by transversal matter currents in the presence of non-standard interactions (NSI). We derived the full oscillation probabilities for two cases (accounting for only those of matter currents and matter currents with NSI) and provided a numerical estimation under astrophysical conditions resembling neutron star mergers. Our analysis incorporates realistic NSI coupling strengths, offering real numerical orders that can be expected in future experiments.

\begin{acknowledgments}
This study was supported by the Russian Science Foundation (project no. 24-12-00084). The work was performed using the scientific infrastructure provided within the scientific program of the National Center for Physics and Mathematics (section no. 8 “Physics of hydrogen isotopes”, project “Fundamental studies in the field of neutrino physics and neutron-rich nuclei using isotopes of hydrogen and helium”).
\end{acknowledgments}

%%%%%%%%%%%%%%%%%%%%%%%%%%%%%%%%
% USE thebibliography
%%%%%%%%%%%%%%%%%%%%%%%%%%%%%%%%


\begin{thebibliography}{}
% book
\bibitem{key1} A.Studenikin, Neutrinos in electromagnetic fields and moving media, Phys. At. Nucl. 67, 993 (2004).

\bibitem{key2} A.Lobanov and A.Studenikin, Neutrino oscillations in moving and polarized matter under the influence of electromagnetic Fields, Phys. Lett. B 515, 94 (2001).

\bibitem{key5} P.Pustoshny and A.Studenikin, Neutrino spin and spin-flavor oscillations in transversal matter currents with standard and nonstandard interactions, Phys. Rev. D 98, 113009 (2018).

\bibitem{keyFarzan} Y.Farzan and M.Tortola, Neutrino oscillations and non-standard interactions, Front. Phys. 6, 10 (2018).

\bibitem{key3} A.Perego et al., Neutrino-driven winds from neutron star merger remnants, Mon. Not. R. Astron. Soc. 443, 3134 (2014).

\bibitem{key4} P.Coloma, C.Gonzalez-Garcia, M.Maltoni, J.Pinheiro, and S.Urrea, Global constraints on non-standard neutrino interactions with quarks and electrons, JHEP 08, 032 (2023).

\bibitem{Grigoriev:2002ew}
A.Grigoriev, A.Lobanov, and A.Studenikin,
%\href{https://doi.org/10.1016/S0370-2693(02)01776-8}
{Effect of matter motion and polarization in neutrino flavour oscillations},
Phys.~Lett.~B \textbf{535}, 187 (2002).
%\href{https://arxiv.org/abs/hep-ph/0202276}{hep-ph/0202276}.

%\cite{Popov:2019nkr}
\bibitem{Popov:2019nkr}
A.Popov and A.Studenikin,
Neutrino eigenstates and flavour, spin and spin-flavour oscillations in a constant magnetic field,
Eur. Phys. J. C \textbf{79} (2019) no.2, 144
%doi:10.1140/epjc/s10052-019-6657-z
%[arXiv:1902.08195 [hep-ph]].
%43 citations counted in INSPIRE as of 23 Jun 2026


\bibitem{key6} A.Studenikin, Electromagnetic neutrinos: The basic interaction processes and constraints from laboratory experiments and astrophysics, Int. J. Mod. Phys. E 33, 2441033 (2024).

	

\end{thebibliography}
\end{document}